\documentclass[aps,prl,twocolumn,amsmath]{revtex4}
\usepackage{graphicx}  
\usepackage{color}  

\begin{document}

\title{Gravitational-wave background from kink-kink collisions on infinite cosmic strings}

\author{Yuka Matsui}
\author{Sachiko Kuroyanagi}
\affiliation{Department of physics and astrophysics, Nagoya University,
Nagoya, 464-8602, Japan, Nagoya University}

\begin{abstract}
  We calculate the power spectrum of the stochastic gravitational-wave
  (GW) background expected from kink-kink collisions on infinite
  cosmic strings.  Intersections in the cosmic string network
  continuously generate kinks, which emit GW bursts by their
  propagation on curved strings as well as by their collisions.
  First, we show that the GW background from kink-kink collisions is
  much larger than the one from propagating kinks at high frequencies
  because of the higher event rate.  We then propose a method to take
  into account the energy loss of the string network by GW emission as
  well as the decrease of the kink number due to the GW backreaction.  We
  find that these effects reduce the amplitude of the GW background at
  high frequencies and produce a flat spectrum. Finally, we obtain a
  constraint on the string tension of $G \mu \lesssim 10^{-5}$ using the
  current upper bound on the GW background by Advanced LIGO, and $G
  \mu \lesssim 4\times 10^{-8}$ using pulsar timing arrays.
\end{abstract}

\maketitle

\section{I. Introduction}
Cosmic strings are one-dimensional topological defects that  may have
been generated during a phase transition in the early Universe
\cite{Kibble:1976sj}.  They are considered to form a network of
infinite strings and loops, both of which have singular structures---called kinks and cusps---which emit strong gravitational-wave (GW)
bursts \cite{Damour:2000wa,Damour:2001bk}. Overlapping bursts form a GW
background, which can be tested by various GW experiments.  In Ref. 
\cite{Kawasaki:2010yi} and our previous work \cite{Matsui:2016xnp},
the power spectrum of the GW background from kinks propagating on
infinite strings was estimated using the number distribution of
kinks derived in Ref. \cite{Copeland:2009dk}.  We have found that the GW
background is generated over a wide range of frequencies, from the
scale of the cosmic microwave background to direct-detection GW
experiments.

In addition to GWs from kink propagation, kink-kink collisions are
also expected to generate a GW background.  Using the kink
distribution derived in \cite{Copeland:2009dk}, in this paper we
calculate the power spectrum of the GW background originating from
overlapping bursts from kink-kink collisions on infinite strings.  We
numerically calculate the kink number distribution as a function of
time and sharpness, which gives the rate of kink-kink collisions, and
estimate the amplitude of the GW spectrum by summing up the
contribution from all of the redshifts.  Since the result shows that
kink-kink collisions generate a large GW background, we take into
account two effects that could modify the estimate of the kink number
distribution due to the large GW emission.  The first effect is the
energy loss of the string network through GW emission, which reduces
the length of infinite strings.  The second is the GW backreaction on
kinks, which smooths out the kink sharpness.  We include these factors in
the calculation of the GW spectrum and, finally, compare it with the
sensitivities of current and future GW experiments and discuss
constraints on the cosmic string tension $G\mu$.

\section{II. Basic equations}
The dynamics of cosmic strings is well described by the Nambu-Goto
equations.  By considering a spatially flat
Friedmann-Lema{\^\i}tre-Robertson-Walker (FLRW) metric, ${\rm d}s^2 =
a^2(\tau) \left (-{\rm d} \tau^2 +{\rm d} {\bf x}^2 \right )$,
choosing the coordinates on the worldsheet as $\tau$ (conformal time)
and $\sigma$ (direction along a cosmic string), and using the gauge
condition $\frac{\partial x^\mu}{\partial \tau}\frac{\partial
  x_\mu}{\partial \sigma}=0$, the Nambu-Goto action gives the
evolution equation
\begin{equation}
  \frac{\partial^2 {\bf x}}{\partial \tau^2} +\frac{2}{a} \frac{{\rm d} a}{{\rm d} \tau} \frac{\partial {\bf x}}{\partial \tau} \Biggl\{1 -\left (\frac{\partial {\bf x}}{\partial \tau} \right )^2 \Biggr\} = \frac{1}{\epsilon} \frac{\partial}{\partial \sigma} \left (\frac{1}{\epsilon} \frac{\partial {\bf x}}{\partial \sigma} \right ) \, , \label{eq:string_EoM}
\end{equation}
where $\epsilon \equiv \sqrt{\frac{(\partial {\bf x}/\partial
    \sigma)^2}{1-(\partial {\bf x}/\partial \tau)^2}}$ is interpreted
as energy per unit $\sigma$.  In Minkowski spacetime, the solution
is given by a linear superposition of left- and right-moving modes,
${\bf x}=({\bf a}+{\bf b})/2$.  Here we introduce new variables ${\bf
  p}_{\pm}$, which represent left- and right-moving modes in the FLRW
spacetime (corresponding to $\partial{\bf a}/\partial\sigma$ and
$\partial{\bf b}/\partial\sigma$ in Minkowski spacetime), as
\begin{equation}
  {\bf p}_{\pm} \equiv \frac{\partial {\bf x}}{\partial \tau} \mp \frac{1}{\epsilon} \frac{\partial {\bf x}}{\partial \sigma} \, . \label{eq:p_pm}
\end{equation}
At a kink, the value of ${\bf p}_{\pm}$ changes discontinuously from
${\bf p}_{\pm, \, 1}$ to ${\bf p}_{\pm, \, 2}$.  We define the
sharpness of the kink as
\begin{equation}
  \psi \equiv \frac{1}{2} (1- {\bf p}_{\pm, \, 1} \cdot {\bf p}_{\pm, \, 2}) \, . \label{eq:psi}
\end{equation}

Cosmic strings obey a scaling law, in which the correlation length
of cosmic strings evolves in proportion to the cosmic time $t$.
The velocity-dependent one-scale (VOS) model \cite{Kibble:1984hp}
gives the evolution equations of the correlation length $L$ and the
velocity $v$ as
\begin{eqnarray}
  \frac{{\rm d}L}{{\rm d}t} & = & HL(1 +v^2) +\frac{1}{2}cpv, \label{eq:L_eq} \\
  \frac{{\rm d}v}{{\rm d}t} & = & (1 -v^2) \left (\frac{k(v)}{L} -2Hv \right ), \label{eq:v_eq}
\end{eqnarray}
where $H$ is the Hubble parameter $H\equiv ({\rm d}a/{\rm d}t)/a$ with
$a(t)$ being the scale factor of the Universe and $k(v)$ is effective
curvature $k(v) = \frac{2 \sqrt{2}}{\pi} \frac{1 -8v^6}{1+8v^6}$.  The
Hubble parameter is calculated as
$H=H_0\sqrt{\Omega_{r}a^{-4}+\Omega_{m}a^{-3}+\Omega_{\Lambda}}$ with
the Hubble constant $H_0=100 h$km/s/Mpc.  We use
$\Omega_{r}h^2=4.31\times 10^{-5}$ and the cosmological parameters
obtained from the {\it Planck} satellite: $h = 0.692$, $\Omega_{\rm m} = 0.308$, 
and $\Omega_{\Lambda}= 0.692$ \cite{Ade:2015xua}.  The second term of
Eq.~\eqref{eq:L_eq} describes the loop production, and the value of
the loop chopping efficiency $c$ is taken as $c \simeq 0.23$
\cite{Martins:2000cs}.  In this paper, we investigate the case of unit
reconnection probability $p=1$, which is typical of field-theoretic
strings.  By simultaneously solving Eqs.~\eqref{eq:L_eq} and
\eqref{eq:v_eq}, we obtain the solution of $L \propto t$ and we define
the coefficient as $\gamma \equiv L/t$.

\begin{figure}
\centering
\includegraphics[width=9cm,clip]{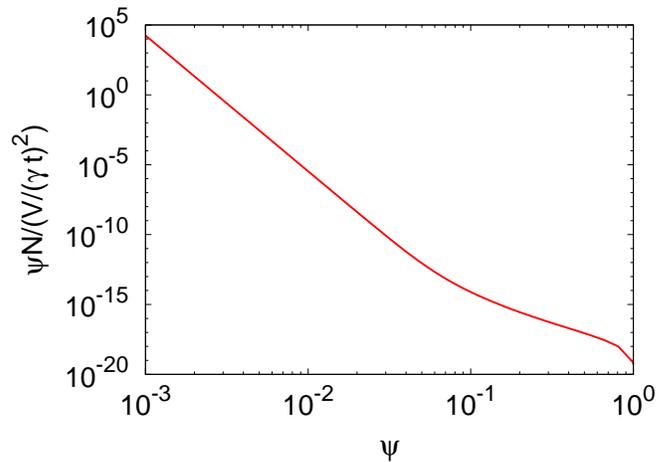}
\caption{The distribution function of kinks on infinite strings.  The
  number of kinks per unit length per logarithmic sharpness is shown
  as a function of sharpness.}
\label{fig:distribution_kink} 
\end{figure}

To calculate the GW background from kinks, we first estimate the
distribution function of kinks $N(\psi,t)$, where $N(\psi,t){\rm
  d}\psi$ is the number of kinks between $\psi$ and $\psi + {\rm
  d}\psi$ within the arbitrary volume $V$.  The evolution equation of
$N(\psi,t)$ is written as \cite{Copeland:2009dk}
\begin{equation}
\frac{\partial N}{\partial t}(\psi, \, t) = \frac{\bar{\Delta} V}{\gamma^4 t^4} g(\psi) +\frac{2 \zeta}{t} \frac{\partial}{\partial \psi} (\psi N(\psi, \, t)) -\frac{\eta}{\gamma t} N(\psi, \, t).  \label{eq:kink_CS_eq}
\end{equation}
The first term is the number of kinks produced by intersecting cosmic
strings.  The second term describes the blunting of kinks due to the
expansion of the Universe.  The third term is the number of kinks lost
into loops.  The parameters of each term in Eq.\eqref{eq:kink_CS_eq}
are described as
\begin{eqnarray}
 \bar{\Delta} & = & \frac{2 \pi}{35} \Bigl\{1 +\frac{2}{3}(1-2v^2) -\frac{1}{11}(1 -2v^2)^2 \Bigr\}, \\
  \zeta & = & (1 -2v^2) \nu,  \\
  \eta & = & \frac{1}{2} c p v,
\end{eqnarray}
where $\nu$ is the parameter describing the time dependence of the scale
factor $a\propto t^\nu$.  By simultaneously solving Eqs.~\eqref{eq:L_eq},
\eqref{eq:v_eq}, and \eqref{eq:kink_CS_eq}, we obtain the distribution
function of kinks.
Figure \ref{fig:distribution_kink} shows the number of kinks on infinite
strings per unit length per logarithmic sharpness $\frac{\psi N(\psi,
  \, t)}{V(t)/(\gamma t)^2}$ as a function of $\psi$.
From the figure, we see that kinks with small sharpness, which are
produced in the radiation-dominated era, are more abundant than the ones with
$\psi\sim 1$, which are produced recently.  A more detailed
explanation is given in Ref. \cite{Matsui:2016xnp}.

\section{III. GW background}
Next, we describe the method to calculate the GW background spectrum
from kink-kink collisions.  For comparison, we also describe the
formalism for kinks propagating on infinite strings, which follows Ref. 
\cite{Matsui:2016xnp}.

The Fourier strain amplitude of GWs from a propagating kink and a
kink-kink collision are given, respectively, by \cite{Binetruy:2009vt}
\begin{eqnarray}
  h_{\rm k}(\psi,f,z) & = & \frac{\psi^{1/2} G\mu  \gamma t}{\{(1+z)f \gamma t \}^{2/3}}\frac{1}{r(z)} \Theta(1-\theta_m), \label{eq:kink_h_eq}  \\
  h_{\rm kk}(\psi,f,z) & = & \frac{\psi G \mu}{(1+z) f} \frac{1}{r(z)} \Theta(1-\theta_m), \label{eq:k-k_col_h_eq}
\end{eqnarray}
where $f$ is the observed frequency so that $(1+z)f$ is the frequency
at the emission, $\mu$ is the cosmic string tension, which becomes
a dimensionless parameter $G\mu$ by multiplying it by the
gravitational constant $G$, $r(z) = \int^z_{0} \frac{{\rm d} z}{H(z)}$
is the distance from the observer, and the first Heaviside step
function $\Theta(1- \theta_m)$ with $\theta_m \equiv \{(1+z)f \gamma t
\}^{-1/3}$ plays the role to cutting off the long-wavelength mode beyond the
string curvature $\gamma t$.

From Eq. (\ref{eq:kink_CS_eq}), we obtain the number of kinks as a
function of sharpness and $\psi \frac{N(\psi, \, t)}{V(t)/(\gamma t)^2}$
gives the number per unit length per logarithmic sharpness.
The inverse of this gives the average interval of kinks with a given
sharpness.  Reference \cite{Kawasaki:2010yi} investigated the GW
background from kinks propagating on infinite strings and found that,
for a given comoving GW frequency $f$, the dominant contribution on the GW
background comes from kinks whose interval is the GW wavelength
$\omega^{-1}$, where $\omega=2\pi f (1+z)$ is the physical GW angular
frequency at the emission redshift $z$.  Using the same discussion
presented in the appendix of Ref. \cite{Kawasaki:2010yi}, we can show that
the same holds for the case of kink-kink collisions.  For the sources
at redshift $z$, the condition is given by
\begin{equation}
  \left (\psi \frac{N(\psi, \, t)}{V(t)/(\gamma t)^2} \right )^{-1} \sim \omega^{-1} \, . \label{eq:psimax}
\end{equation}
From now on, we denote the
sharpness satisfying Eq.~\eqref{eq:psimax} as $\psi_{\rm m}$.

The power spectrum of the GW background is often characterized by
$\Omega_{\rm gw}\equiv ({\rm d}\rho_{\rm GW}/{\rm d}{\rm ln}f)/\rho_c
$, where $\rho_{\rm GW}$ is the energy density of GWs and
$\rho_c\equiv 3H^2/8\pi G$ is the critical density of the Universe.
Using the effective GW burst rate $n(f,z)\equiv \frac{1}{f}\frac{{\rm
    d}\dot{N}}{{\rm d}\ln z}$, where $\dot{N}(f,z)$ is the event rate
of GW bursts with frequency $f$ at redshift $z$, the power spectrum
today is given by integrating contributions from all redshifts,
\begin{eqnarray}
  \Omega_{\rm GW}(t_0,f) = \frac{2\pi^2f^2}{3H_0^2} \int \frac{{\rm d}z}{z}\Theta(n(\psi_{\rm m},f,z)-1)\nonumber\\
  \times n(\psi_{\rm m},f,z)h^2(\psi_{\rm m},f,z), \label{eq:Omega_gw}
\end{eqnarray}
where we included the step function $\Theta(n(f,z)-1)$ to exclude
rare bursts that do not overlap enough to form a GW background.
Note that we only consider contributions from $\psi_{\rm m}$
according to the discussion around Eq.~\eqref{eq:psimax}.

When one only considers contributions from kinks between $\ln\psi_{\rm
  m}$ and $\ln\psi_{\rm m}+{\rm d} \ln\psi_{\rm m}$, the event rate
$n(\psi_{\rm m},f,z)$ can be estimated as
\begin{eqnarray}
  && n(\psi_{\rm m},f,z)  =  \frac{1}{f}
  \times \left (\text{rate of GW bursts per kink}) \right . \nonumber \\
  && \times (\text{\# of kinks per unit volume } \frac{\psi_{\rm m} N(\psi_{\rm m},t)}{V}) \times \frac{{\rm d}V(z)}{{\rm d} \, {\rm ln} z}, \nonumber \\
\end{eqnarray}
where $\frac{{\rm d} V}{{\rm d} z}=\frac{1}{z}\frac{{\rm d} V}{{\rm d}
  \ln z} = \frac{4\pi a^3r^2(z)}{H(z)}$ is the volume between $z$ and
$z+{\rm d}z$.  Using the beaming angle of GWs, $\frac{\theta_m}{2}$,
and the typical curvature of the string $\gamma t$, the rate of GW bursts
from a propagating kink is given by $\frac{\theta_m}{2(1+z)\gamma t}$,
where $(1+z)$ is added to take into account the redshift of the time
interval.  For kink-kink collisions, the number of kinks per unit time crossing the path of any given kink is given by $\frac{\psi_{\rm m}
  N(\psi_{\rm m},t)}{V/(\gamma t)^2}$.  By multiplying by $\frac{1}{2}$
to avoid double counting and taking into account the redshift, the
rate of GW bursts per kink is given by $\frac{(\gamma
  t)^2}{2(1+z)}\frac{\psi_{\rm m} N(\psi_{\rm m},t)}{V}$.  Note that a
kink-kink collision emits GWs in all directions, and thus we do not
multiply the rate by a beaming angle.  In summary, the burst rates for propagating
kinks and kink-kink collisions are written, respectively, as
\begin{eqnarray}
  n_{\rm k}(\psi_{\rm m},f,z) & = & \frac{1}{f} \frac{\theta_m}{2(1+z)\gamma t} \frac{\psi_{\rm m} N(\psi_{\rm m},t)}{V} \frac{{\rm d}V(z)}{{\rm d} \, {\rm ln} z}, 
  \label{eq:nk}\\
   n_{\rm kk}(\psi_{\rm m},f,z) & = & \frac{1}{f} \frac{(\gamma t)^2}{2(1+z)} \Bigl\{\frac{\psi_{\rm m} N(\psi_{\rm m},t)}{V} \Bigr\}^2  \frac{{\rm d}V(z)}{{\rm d} \, {\rm ln} z}. \nonumber
   \label{eq:nkk} \\
\end{eqnarray}
By substituting Eqs.~\eqref{eq:k-k_col_h_eq} and \eqref{eq:nkk} [Eqs.~\eqref{eq:kink_h_eq} and \eqref{eq:nk}]
into Eq.~\eqref{eq:Omega_gw}, we obtain the GW background spectrum for
kink-kink collisions (propagating kinks).  Note that the
value of $\psi_{\rm m}$ is time dependent and is estimated using
Eq.~\eqref{eq:psimax} at every time step of the numerical calculation.

\begin{figure}
\centering
\includegraphics[width=9cm,clip]{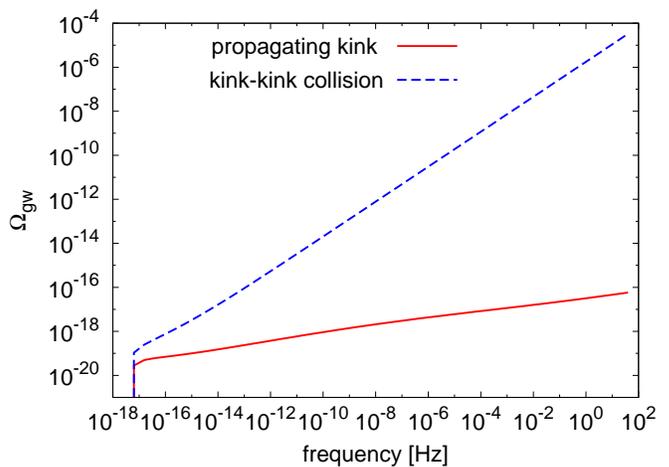}
\caption{The power spectrum of the GW background from 
propagating kinks (red solid) and kink-kink collisions (blue dashed). 
  For both lines, we assume $G\mu = 10^{-11}$.}
\label{fig:GW_kink_and_k-k_col} 
\end{figure}

In Fig.~\ref{fig:GW_kink_and_k-k_col} we compare the power spectrum
of the GW background from propagating kinks and kink-kink collisions.
Given the fact that $\psi \frac{N(\psi, \, t)}{V(t)/(\gamma t)^2} $ is
a decreasing function of $\psi$, Eq.~\eqref{eq:psimax} indicates that
the high-frequency GWs are produced by kinks with small sharpness,
which have a high event rate.  The large amplitude difference at high
frequencies between the two cases arises because the event rate of
kink-kink collisions increases in proportion to the square of the kink
number, while the dependence is linear in the case of propagating
kinks.  
One finds that the overproduction of GWs at high frequencies violates the constraints from big bang nucleosynthesis and the cosmic microwave background, $\Omega_{\rm GW}\lesssim 2 \times 10^{-6}$
\cite{Cabass:2015jwe}. However, this is not the final result and, in
fact, it will be solved in the next section.

\section{IV. Effects of GW emission}
As shown in the previous section, the power spectrum increases
dramatically towards high frequencies in the case of kink-kink
collisions.  One may be concerned that a large amount of GW emissions could
change the number of infinite strings, since the energy of the string
network is transferred to GWs.  In addition, the backreaction of GW emission could
smooth out the sharpness of kinks and reduce the power of GW emission.
In this section, we take these two effects into account by modifying
the VOS equation [Eq.~\eqref{eq:L_eq}] and the evolution equation of the
kink distribution [Eq.~\eqref{eq:kink_CS_eq}], and recalculating the
GW power spectrum.

Let us first consider the effect of GW radiation on the VOS equation
\cite{Austin:1993rg,Copeland:1999gn}.  The energy of GW emission from
one kink-kink collision is estimated as $E_{\rm GW}\sim 2 \pi^3 \psi^2
G\mu^2\omega^{-1}$ \cite{Binetruy:2009vt}.  Here, the factor $2 \pi^3$
is added to make $E_{\rm GW}$ consistent with the expression for
$\Omega_{\rm GW}$, that is, with the choice of the factor in
front of $h_{\rm kk}$ in Eq. \eqref{eq:k-k_col_h_eq}.  Considering the
energy conservation law for the string network density
$\rho_{\infty}=\frac{\mu}{L^2}$ \cite{Martins:2000cs}, the loss of
energy density as GW radiation is given by
\begin{eqnarray}
    \frac{{\rm d} \rho_{\infty}}{{\rm d}t}
    & = & - \int_{0}^{1} {\rm d}\psi_{\rm m} \, E_{\rm GW}\times
    \text{(\# of GWs \nonumber} \\
    && \hspace{70pt} \text{ per unit volume, time, d$\psi_m$)}, \nonumber \\
    & = & - \int_{0}^{1} {\rm d}\psi_{\rm m} \, 2 \pi^3
    \psi_{\rm m}^2 G\mu^2\omega^{-1} ~
    \frac{\psi_{\rm m}}{2}
    \left\{\frac{N(\psi_{\rm m},t)}{V/(\gamma t)^2}\right\}^2\frac{1}{(\gamma t)^2} .\nonumber\\
\end{eqnarray}
  Here, the integral in terms of d$\psi_{\rm m}$ corresponds to taking
  into account GWs of all frequencies.  By rewriting
  $\rho_{\infty}$ in terms of $L$ and adding it to
  Eq.~\eqref{eq:L_eq}, we get
  \begin{equation}
    \frac{{\rm d}L}{{\rm d}t} = HL(1 +v^2) +\frac{1}{2}cpv 
    +\frac{\pi^3 G \mu}{2} \gamma t \int_{0}^{1} {\rm d}\psi_{\rm m} \, \frac{N(\psi_{\rm m},t)}{V/(\gamma t)^2} \psi_{\rm m}^2 , \label{eq:L_eq_with_GWR}
  \end{equation}
  where we have used Eq.~\eqref{eq:psimax} to replace $\omega$.

Next, we consider the GW backreaction on kinks and estimate the effect on the
kink distribution.  Before presenting the equations, let us compare
the energy of one kink and the GW energy at one collision.  When we treat
a kink as a small perturbation $\delta {\bf p}_\pm$
\cite{Siemens:2001dx,Copeland:2006if}, the energy of the kink is
estimated as $E_{\rm kink}=\mu (\delta {\bf p}_\pm)^2\Delta \ell \sim
\mu\psi \Delta \ell$ for a given length $\Delta \ell$, where we have
used Eq.~\eqref{eq:psi} in the second step.  From
Eq.~\eqref{eq:psimax}, we expect that kinks contributing to the GW
background are distributed with an average interval of $\omega^{-1}$, so we take $\Delta\ell\sim \omega^{-1}$.  By taking the ratio $E_{\rm
  GW}/E_{\rm kink}=2 \pi^3 \psi G\mu$, we find that the fraction of energy
going to GW emission is initially as small as $\sim G\mu$ for newly
formed kinks $\psi\sim 1$, and the fraction gets even smaller when the
kink sharpness is made smaller by the expansion of the Universe. Thus,
when we consider the GW energy at one collision, the GW backreaction
seems to be negligible.

However, the accumulation of a small GW backreaction through a huge number
of collisions could change the kink distribution. This can be
implemented as a modification of Eq. \eqref{eq:kink_CS_eq}.  By
considering the energy fraction going to GWs, the backreaction term
can be written as
\begin{eqnarray}
   && (\text{\# of kinks lost by GW emission per V, time, d$\psi_m$})  \nonumber \\
  && \sim \frac{E_{\rm GW}\times(\text{\# of GWs per V, time, d$\psi_m$})}
         {E_{\rm kink}\times(\text{\# of kinks per V})}  \nonumber \\
  && \hspace{140pt}
  \times (\text{\# of kinks per V})  \nonumber \\
  && \sim (2 \pi^3 \psi G \mu) \frac{\frac{1}{2} \psi \Bigl\{\frac{N(\psi)}{V/(\gamma t)^2} \Bigr\}^2 \frac{V}{(\gamma t)^2}}{N(\psi, \, t)} N(\psi, \, t) .
\end{eqnarray}
By adding this term, Eq.~\eqref{eq:kink_CS_eq} becomes
\begin{eqnarray}
  \frac{\partial N}{\partial t}(\psi, \, t) & = & \frac{\bar{\Delta} V}{\gamma^4 t^4} g(\psi) +\frac{2 \zeta}{t} \frac{\partial}{\partial \psi} (\psi N(\psi, \, t)) -\frac{\eta}{\gamma t} N(\psi, \, t) \nonumber \\
  & \, & \hspace{50pt} -\frac{\pi^3 G \mu \psi^2(\gamma t)^2}{V} N^2(\psi, \, t).
  \label{eq:kink_CS_eq_with_BR}
\end{eqnarray}

\begin{figure}
\centering
\includegraphics[width=9cm,clip]{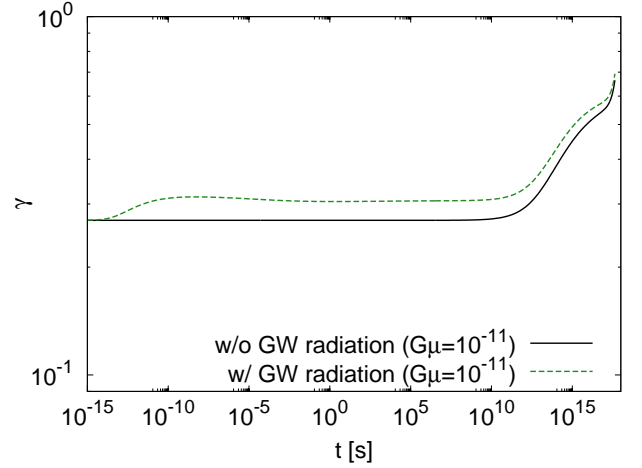}
\caption{The time evolution of $\gamma$ calculated using
  Eqs.~\eqref{eq:L_eq_with_GWR} and \eqref{eq:kink_CS_eq_with_BR} for
  $10^{-11}$.  For comparison, we also show the
  line calculated without the GW radiation term. }
\label{fig:gamma_GWR} 
\end{figure}

\begin{figure}
\centering
\includegraphics[width=9cm,clip]{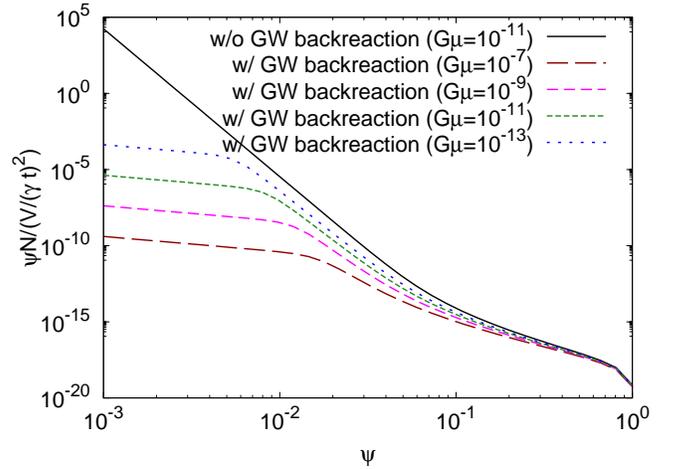}
\caption{The distribution function of kinks calculated using
    Eqs.~\eqref{eq:L_eq_with_GWR} and \eqref{eq:kink_CS_eq_with_BR}.
  The number of kinks per unit length per logarithmic sharpness is
  shown as a function of sharpness.  Each line represents
    a different tension, from $G \mu = 10^{-7}$ to $10^{-13}$.  For
    comparison, we also show the line calculated using
    Eqs.~\eqref{eq:L_eq} and \eqref{eq:kink_CS_eq}. }
\label{fig:distribution_kink_BR} 
\end{figure}

In Figs.~\ref{fig:gamma_GWR} and \ref{fig:distribution_kink_BR}, we
show the time evolution of $\gamma$ and the kink distribution,
respectively, calculated by simultaneously solving the VOS equations
with GW radiation [Eqs.~\eqref{eq:L_eq_with_GWR} and \eqref{eq:v_eq}]
and the equation for the kink distribution with GW backreaction [Eq.~\eqref{eq:kink_CS_eq_with_BR}].  
From Fig.~\ref{fig:gamma_GWR}, we
see that the correlation length does not change at first, but starts
to increase when the GW radiation term becomes non-negligible compared
to the Hubble term.
In Fig.~\ref{fig:distribution_kink_BR}, we find that the number of kinks
with small sharpness is suppressed, since the backreaction term in
Eq.~\eqref{eq:kink_CS_eq_with_BR} affects the distribution when $N$ is
large as it has a $\propto N^2$ dependence.  We also see that the
effect extends to larger sharpness when $G\mu$ is larger.  The slope
of the distribution function becomes gentler for large $G\mu$ because the
value of $\gamma$ is larger due to the modification in
Eq.~\eqref{eq:L_eq_with_GWR}.

\begin{figure}
\centering
\includegraphics[width=9cm,clip]{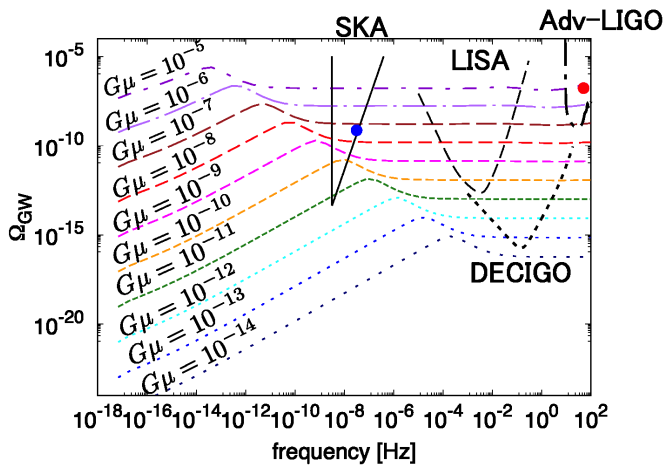}
\caption{The power spectra of the GW background from kink-kink
  collisions for different string tensions, from $G\mu = 10^{-5}$ to$
  10^{-14}$.  We also show the sensitivities of the future
  observational instruments SKA, LISA, DECIGO, and Advanced LIGO. The
  red and blue circles represent the current upper limit on the GW amplitude
  by Advanced LIGO and NANOGrav.}
\label{fig:GW_k-k_col_diff_mu} 
\end{figure}

Finally, in Fig.~\ref{fig:GW_k-k_col_diff_mu}, we plot the power
spectra of the GW background from kink-kink collisions for different
$G\mu$.  We see that high-frequency GWs are suppressed when we use
the kink distribution with the GW modification.  This is mainly
because the number of small kinks is suppressed by the GW
backreaction term in Eq.~\eqref{eq:kink_CS_eq_with_BR}.  We find that the
suppression takes place at late times and it occurs earlier for
smaller sharpness, which corresponds to high-frequency GWs.  As a
result, GWs of the high-frequency plateau are dominantly produced in 
kink-kink collisions in the radiation-dominated era, while ones in the
small bump are produced in the matter-dominated era and ones in the
low-frequency slope are generated today without being affected by the
suppression.  In the figure, we compare the spectra with the sensitivity
curves of the future GW experiments SKA, LISA, DECIGO,
and Advanced LIGO.
We also plot the current upper limit on the GW background amplitude from the 
first observing run of Advanced LIGO, $\Omega_{\rm GW} < 1.7 \times
10^{-7}$ at 20-86 Hz \cite{TheLIGOScientific:2016dpb,Abbott:2017mem}, and the 11-year
data set of NANOGrav, $\Omega_{\rm GW} h^2 < 3.4 \times 10^{-10}$ at
$3.2 \times 10^{-8}$ Hz \cite{Arzoumanian:2018saf}.  We find that the
current Advanced LIGO upper limit gives a constraint on the string
tension of $G \mu \lesssim 10^{-5}$, and the NANOGrav constraint gives
$G \mu \lesssim 4\times 10^{-8}$.  In the future, Advanced LIGO operating at its  full
design sensitivity could provide $G \mu \lesssim 10^{-7}$, and pulsar
timing with SKA could reach $G \mu \sim 10^{-11}$.  With satellite
experiments, we may be able to reach $G \mu \sim 10^{-11}$ using LISA
and $G \mu \sim 10^{-13}$ using DECIGO.

\section{V. Discussion}
Let us first discuss the spectral dependence of the GW
  background spectrum.  When the GW backreaction is absent, one can
  find from Fig. \ref{fig:GW_kink_and_k-k_col} that the GW spectrum
  from kink-kink collisions scales as $\Omega_{\rm GW}
  \propto f^{0.77}$. This dependence is explained as follows.
  Substituting Eqs.~\eqref{eq:k-k_col_h_eq} and \eqref{eq:nkk} into
  Eq.~\eqref{eq:Omega_gw}, replacing the number of kinks with $f$
  using Eq.~\eqref{eq:psimax}, and leaving only the frequency and 
  time dependence, we obtain
\begin{equation}
  \Omega_{\rm GW}
  \propto
  \int {\rm d} ({\rm ln}t) \, \frac{\psi_{\rm m}^2}{t(1+z)^3}f.
  \label{eq:OmegaGWapprox}
\end{equation}
In our numerical calculation without the GW backreaction, we find that the
contribution to the integration of $\Omega_{\rm GW}$ gets larger as
the time increases for all of the frequencies.  Thus, the shape of the GW
spectrum is determined by the kink distribution today.  From
Fig.~\ref{fig:distribution_kink}, we find $\psi \frac{N(\psi, \,
  t)}{V(t)/(\gamma t)^2} \propto \psi^{-8.8}$, and we get $\psi_{\rm
  m} \propto f^{-1/8.8}$ using Eq.~\eqref{eq:psimax}. Substituting
this into Eq. \eqref{eq:OmegaGWapprox}, we get $\Omega_{\rm GW}
\propto f^{0.77}$.  This frequency dependence continues up to the
frequency where the oldest kinks (which have smallest sharpness) can
generate GWs. Higher-frequency GWs are generated by kinks with smaller
sharpness and the amplitude of the GW background starts to decrease at
the frequency corresponding to the smallest kinks. This frequency is
determined by 
the moment of time when cosmic strings were generated, which strongly
depends on the generation model. Thus, in this paper we do not
discuss the high-frequency behavior around the cutoff frequency.

\begin{figure*}
\begin{minipage}{0.48\hsize}
\begin{center}
      \includegraphics[width=9cm,clip]{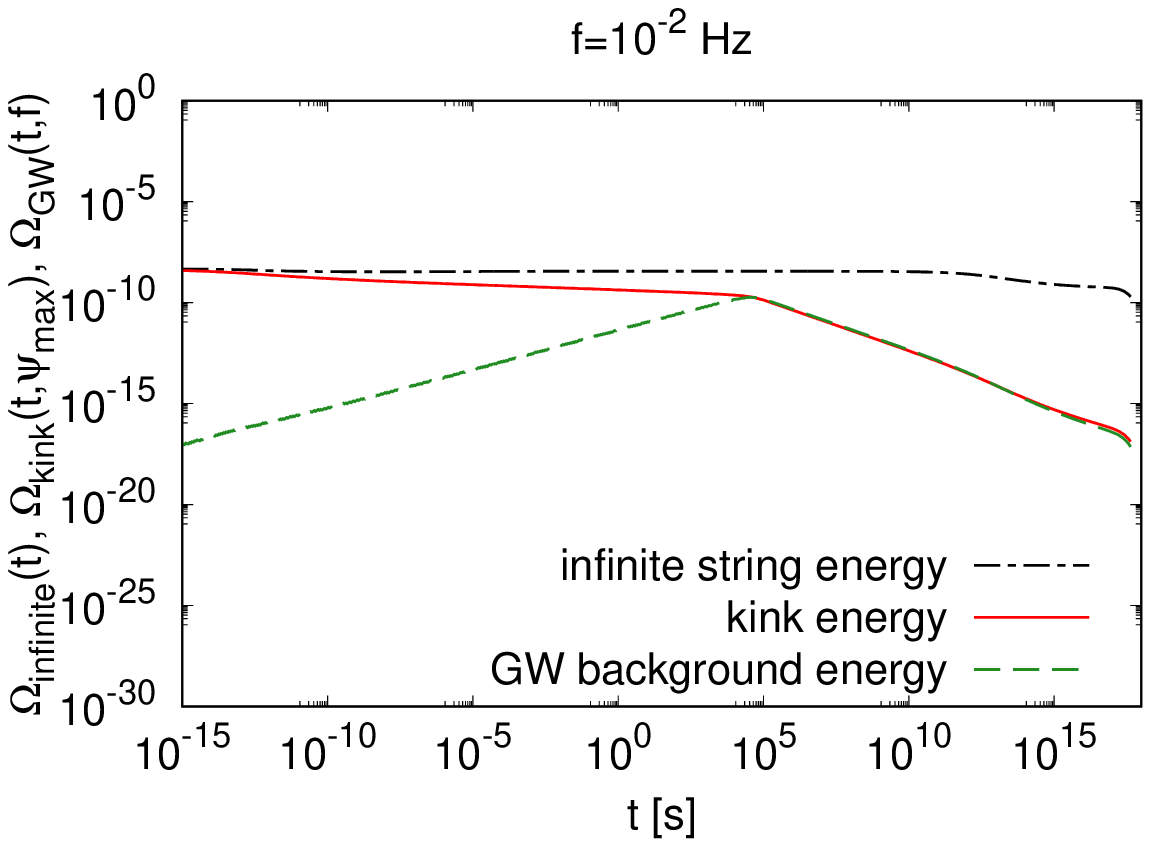}
    \end{center}
  \end{minipage}
  \begin{minipage}{0.48\hsize}
    \begin{center}
     \includegraphics[width=9cm,clip]{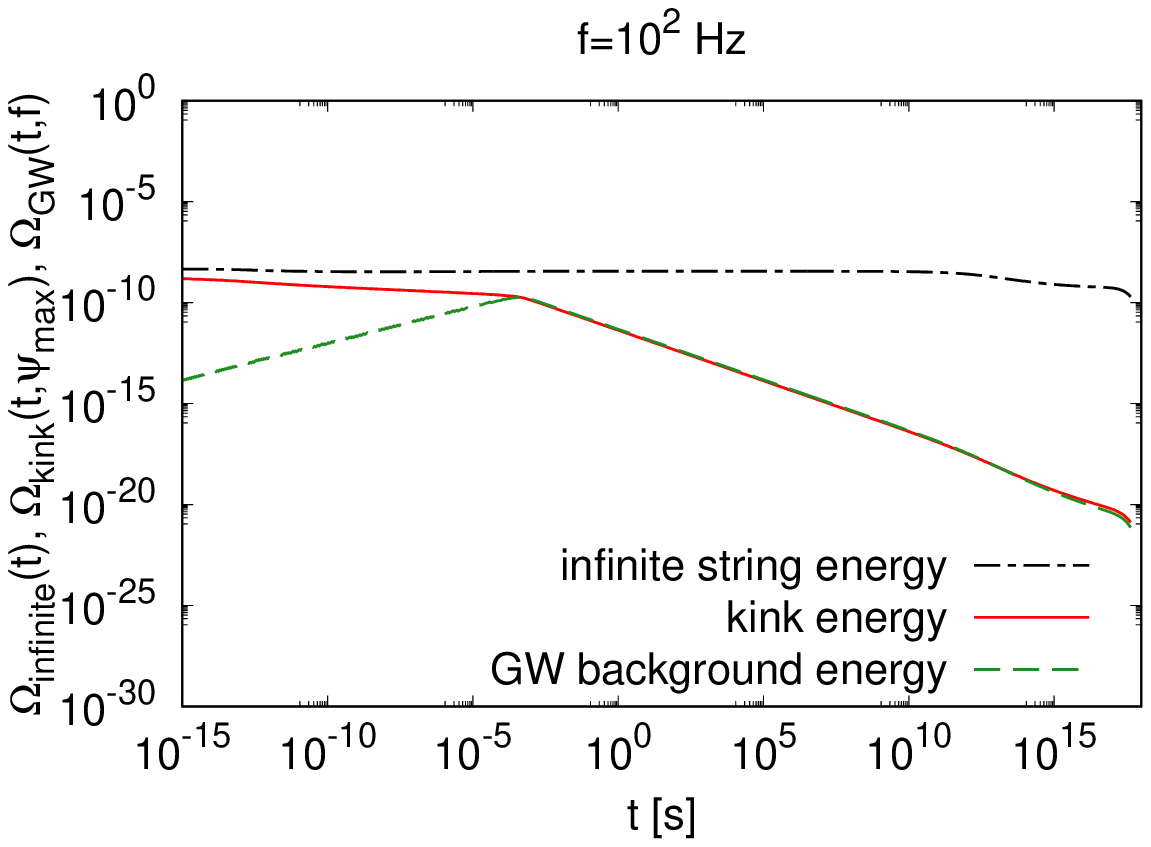}
    \end{center}
  \end{minipage}
  \caption{The time evolution of the infinite string energy density, the kink energy density, and the GW energy density produced at time $t$. The left panel is for kinks contributing to a GW frequency $f = 10^{-2}$Hz and the right panel is for $10^2$Hz.}
\label{fig:time_evo_kink_energy_and_GW_background}
\end{figure*}

When we take the GW backreaction into account, we find that the number
of kinks with small sharpness is suppressed, which reduces
high-frequency GWs and creates a flat plateau in the spectrum.  The 
reason for the flat spectrum is the following.  The GW backreaction
starts to affect kinks with small sharpness first, and the effect
gradually extends to larger sharpness.  Let us define the transition
sharpness as $\psi_{\rm m,cut}(t)$, below which kinks are affected by
the GW backreaction at time $t$.  In the numerical calculation, we
find that the contribution to the integration of $\Omega_{\rm GW}$
peaks when the backreaction starts to take affect, namely when $\psi_{\rm
  m}(t) = \psi_{\rm m,cut}(t)$.  So let us evaluate
Eq. \eqref{eq:OmegaGWapprox} at the time $t_c$, which satisfies
$\psi_{\rm m}(t_c) = \psi_{\rm m,cut}(t_c)$. Here we focus on the
radiation-dominated era since $t_c$ is typically before 
radiation-matter equality for high-frequency GWs.  Using $t \propto
\frac{1}{(1+z)^2}$ and taking out only the contribution at $t_c$,
Eq. \eqref{eq:OmegaGWapprox} becomes
\begin{equation}
  \Omega_{\rm GW} \propto \frac{\psi_{\rm m,cut}^2(t_c)}{1+z_c}f.
  \label{OmegaGWapprox2}
\end{equation}
Here, $z_c$ is the redshift at $t=t_c$, which depends on the
frequency of interest $f$.  Let us first see the time dependence of
$\psi_{\rm m,cut}$.  The GW backreaction starts to take effect when the
fourth term becomes larger than the second and third terms in
Eq. \eqref{eq:kink_CS_eq_with_BR}. Thus, we have
\begin{equation}
  \left(\frac{\eta}{\gamma} -2\zeta \right ) \frac{N}{t_c} = \pi^3 G \mu
  \psi_{\rm m,cut} \Bigl\{\psi_{\rm m,cut} \frac{N}{V/(\gamma t_c)^2} \Bigr\}N.
  \label{eq:N_2_3=backreaction}
\end{equation}
At early times, the backreaction term is negligible and the kink
number evolves as $\psi \frac{N(\psi)}{V/(\gamma t)^2} \propto
t^{-1}$, which is the analytic solution of Eq. \eqref{eq:kink_CS_eq}
detailed in Ref. \cite{Copeland:2009dk}. By substituting this into Eq.\eqref
{eq:N_2_3=backreaction}, we find
that $\psi_{\rm m,cut}$ does not depend on time in the
radiation-dominated era. The relation between $z_c$ and $f$ can be
obtained using Eq.~\eqref{eq:psimax} as $2\pi f =\psi \frac{N(\psi, \,
  t)}{V(t)/(\gamma t)^2} \frac{1}{1+z} \propto (1+z)$. Applying this
relation to Eq. \eqref{OmegaGWapprox2}, we get $\Omega_{\rm GW}
\propto f^{0}$.

Next, let us see how the energy of GWs is balanced in the string network.
We define the energy density parameter of kinks as  
\begin{eqnarray}
  \Omega_{\rm kink} (t, \psi_{\rm m})
  &=& \frac{E_{\rm kink}\times (\text{\# of kinks per unit volume})}{\rho_c}\nonumber\\
  &\sim& \frac{\mu\psi_{\rm m}\omega^{-1}\cdot \psi_{\rm m}\frac{N}{V/(\gamma t)^2}/(\gamma t)^2}{3H^2/(8\pi G)}\\
  &\sim& \frac{8 \pi G \mu}{3 \gamma^2 t^2 H^2} \psi_{\rm m}.
\end{eqnarray}
In the second step, we have used Eq. \eqref{eq:psimax}. Note that this can be written as
  $\Omega_{\rm kink}=\psi_m\Omega_{\rm infinite}$, where $\Omega_{\rm infinite}\equiv\rho_\infty/\rho_c$. This
  indicates that the kink
energy is always smaller than the total energy density of infinite
strings by the order of the sharpness $\psi_{\rm m}$.  In
Fig. \ref{fig:time_evo_kink_energy_and_GW_background} we plot the
time evolution of
the infinite string energy density, 
the kink energy density, and the GW energy
density produced at time $t$ [the integrand of Eq. \eqref{eq:Omega_gw}
before redshifting]. The two panels show the GW frequencies $f=10^{-2}$ and $10^2$Hz, which correspond to different values of $\psi_{\rm m}$.
As one can see, the energy
of GWs increases at the beginning, and when it becomes comparable to
the kink energy both the kink and GW energies start to decrease and
evolve together.  This behavior is due to the fact that the GW
energy is balanced by the kink energy thanks to the GW terms added to
the VOS equation [Eq. \eqref{eq:L_eq_with_GWR}] and the evolution
equation for kink number density [Eq. \eqref{eq:kink_CS_eq_with_BR}].
In summary, we find that the kink and GW energies become of the same
order when the GW terms are turned on, and they always stay below the total
energy of the scaling string network by the order of $\psi_{\rm m}$.

Finally, let us comment on previous works. The GW spectrum from small
structures on infinite strings was calculated analytically in
Refs. \cite{Sakellariadou:1990ne,Hindmarsh:1990xi} and numerical
simulations for GWs from infinite strings were performed in
Ref. \cite{Figueroa:2012kw}. They all predicted a smaller GW amplitude
compared to our result. We believe that the reason is because those
previous studies only considered kinks with large sharpness $\sim 1$
(for simplicity in the analytic study, and because of the resolution in
the simulation study), while our method based on solving the evolution
equation of kink distribution (established in
Ref. \cite{Copeland:2009dk}) enables us to take into account kinks with
much smaller sharpness. In fact, we have seen that the enhancement of
GWs occurs at high frequencies, which are mainly produced by kinks
with small sharpness.

GWs from kink-kink collisions on loops were considered in Refs.
\cite{Binetruy:2010cc,Ringeval:2017eww,Jenkins:2018nty}.  Although
their estimate has some uncertainty since the number of kinks on one
loop was taken as a free parameter in their calculation, it has been
shown that a large GW background can be expected from kink-kink
collisions on loops. We would like to mention that our estimate for the
kink number distribution may help to determine the exact number of loops
and may provide more concrete predictions.

\section{VI. Conclusion}
There have been many efforts to search for and constrain cosmic
strings with cosmological observations.  In this paper, we have shown
a new way to test the existence of cosmic strings by
considering kink-kink collisions on infinite strings.  We have
presented formulas to calculate the GW power spectrum from kink-kink
collisions, which predict a much larger GW amplitude compared to the one
from kink propagation.  Furthermore, we have investigated the effect
of GW radiation and backreaction on the scaling behavior and kink
distribution, and found that these effects reduce the GW amplitude at
high frequencies. Finally, by comparing with the upper bounds on the GW
background amplitude from ongoing experiments, we obtained constraints
on the string tension of $G \mu \lesssim 10^{-5}$ from Advanced LIGO, and
$G \mu \lesssim 4\times 10^{-8}$ from NANOGrav.  Although the current
pulsar timing constraint from loops is stronger than these bounds, we
would like to stress that our prediction based on infinite strings
does not have any ambiguity on the initial loop size, which has been
under debate and could weaken the pulsar timing constraint from loops.
Therefore, our result can be used as an independent test of cosmic
strings.

\section{Acknowledgments}
  This work is partially supported by the Grant-in-Aid for Scientific
  Research from JSPS, Grant Number 17K14282, and by the Career
  Development Project for Researchers of Allied Universities (S. K.).  We
  are deeply grateful to Dani\`ele Steer for providing the initial
  idea and for useful comments which helped to enhance the quality
  of the work.  S. K. would like to thank Jose J. Blanco-Pillado and
  Teruaki Suyama for helpful discussion.

\end{document}